\documentclass[reqno]{amsproc}

\usepackage[final]{epsfig}

 \usepackage{epstopdf}
\newtheorem{theorem}{Theorem}[section]
\newtheorem{lemma}[theorem]{Lemma}

\theoremstyle{definition}
\newtheorem{definition}[theorem]{Definition}

\theoremstyle{remark}

\numberwithin{equation}{section}

\begin{document}
\title[Stability of absolutely continuous spectra]{Fluctuation based proof of the stability of ac spectra of random operators on tree graphs}

\author{Michael Aizenman} %
             \address{Departments of Mathematics and Physics,
            Princeton University, Princeton NJ 08544, USA }
             \email{aizenman@princeton.edu}% 
 
\author{Robert Sims}
\address{Department of Mathematics, University of California at Davis,
Davis CA 95616, USA} \email{rjsims@math.ucdavis.edu}%  
 
\author{Simone Warzel} %
\address{Department of Mathematics,
            Princeton University, Princeton NJ 08544, USA}%
\email{swarzel@princeton.edu}%
 
\thanks{\noindent The article covers talks given at  
UAB Int. Conf. on Diff. Eq. and Math. Phys., Birmingham, March 2005 (RS and SW), and at AMS Snowbird Conference, Utah, June 2005 (MA)} 

\date{January 1, 1994 and, in revised form, June 22, 1994.}


\begin{abstract} 
We summarize recent works on the stability 
under disorder of the absolutely continuous spectra of random
operators on tree graphs.  The cases covered include:
Schr\"odinger operators with random potential, quantum graph
operators for trees with randomized edge lengths, and
radial quasi-periodic operators perturbed by  random potentials.
\end{abstract}

\maketitle

\section{Introduction}

The incorporation of disorder is known to have strong effects on the spectral properties of self adjoint operators and the dynamics they generate.  The phenomenon is relevant for models of physics, where it carries significant implications for the conduction properties of metals, the quantum hall effect, and the properties of quantum networks.  
As has been shown in various disciplines,  disorder is a fruitful subject  for mathematics.   Yet, our understanding of its implications  for random operators  is still  underdeveloped.   This is exemplified by the fact that there is, at present, no example of a local operator in a finite dimensional 
space for which the existence of continuous spectrum has been  established in the presence of weak  but extensive, i.e., homogeneously 
random, disorder.   
Until recently, the one case for which 
there has been a constructive result, due to A.~Klein~\cite{Klein95,Klein98}, 
is the discrete Schr\"odinger operator with an iid random 
potential on a regular Bethe lattice.  Our recent work presents another method, with a different range of applicability, for the proof of the stability, under weak disorder, of ac spectra of tree operators.  In this article we  summarize the results and the main ideas which play a role in their derivation. 

\subsection{Examples of operators with disorder} 

A guiding example of a local generator of quantum dynamics is  
the Schr\"odinger operator 
\begin{equation} 
H_0 = -\Delta +U 
\end{equation} 
 acting in  $L^2({\mathbb R} ^d)$ with $\Delta$ the Laplacian and $U$ a potential which, for instance,
 could be periodic.   Electric and magnetic fields can be 
 taken into account as well, and in the latter case the operator loses 
 its time reversal symmetry -- the operator, and its eigenfunctions, are no longer real.  
 Disorder can be incorporated through the addition of a random potential, e.g.,  
\begin{equation} 
V(\omega) = \sum_{\alpha \in {\mathbb Z}^d} \omega_{\alpha} u(\cdot-\alpha) \, , 
\end{equation} 
 with 
$\{ \omega_{\alpha} \}_{\alpha}$ iid random variables and $u(\cdot)$ a local 
`potential bump'.  In this fashion one obtains a random operator: 
\begin{equation}
H_{\lambda}(\omega) \ = \ H_0 + \lambda \, V(\omega)  \, .
\end{equation}

The issues of interest can also be studied in the simpler context of 
the discrete analog of $H_{\lambda}(\omega) $  acting in $\ell^2({\mathbb G})$, 
where ${\mathbb G}$ is a graph, e.g.\ ${\mathbb Z}^d$, and the random potential 
is described by a collection of random variables associated with the graph vertices. 

\begin{figure}[ht]
    \begin{center}
    \leavevmode
    \includegraphics[width=2.4in]{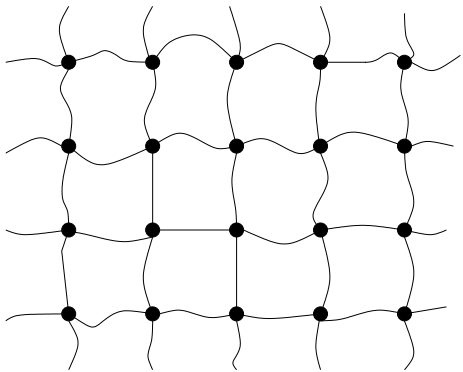} 
  \caption{A mesh with random length on which the quantum graph operator $- d^2/dx^2$  
  may be defined.} 
\label{micro}
\end{center}
\end{figure}

Another example, which provides an interesting test-case for a number of 
concepts -- effects of randomness, trace formulae, quantum wires -- 
is the quantum graph operator $H_0 = - d^2/dx^2$  
acting on 
functions supported on a `mesh of wires'. 
Accordingly,  $H_0 $ is defined in  $ \bigoplus_{e \in E} L^2(0, L_{e}) $, where $e$ ranges over the set of bonds of a graph, and 
$L_{e}$ is the length of the corresponding edge.  It 
is made self-adjoint through a choice of boundary 
conditions at the graph vertices. 
A natural way to incorporate disorder is to vary the edge lengths, 
e.g., letting it be of the form 
\begin{equation}
L_{e}(\omega) \ = \ L_0 \, e^{\lambda \omega_{e} }
\end{equation}
with $\{\omega_{e}\}$ iid random variables, as depicted in Figure~\ref{micro}.    Another form is through random vertex rules, as in~\cite{KS99}.

The disorder is termed here extensive if its distribution is homogeneous.  It is convenient to note that the  
distribution can be ergodic under the symmetry group of the (lattice) translations 
or graph isomorphism. 
This property guarantees \cite{KS,Pas80,KirMar82a} that the spectra and 
spectral components of such random operators are deterministic
in the sense that there exist Borel sets 
$ \Sigma, \Sigma_{ \rm ac}, \Sigma_{ \rm pp}, \Sigma_{ \rm sc},$ such that with probability one, i.e., almost surely, the spectrum of $H(\omega)$ is given by $\Sigma$:  
\begin{equation}
	\sigma(H(\omega)) \ \stackrel{a.s.}{=} \ \Sigma \, ,
\end{equation}
and similarly for the pure-point, the absolutely continuous, and the singular-conti\-nuous 
components of the spectra.  

\subsection{Spectral regimes} 

The {\em absolutely continuous} (ac) component of the spectrum is of particular interest  in condensed matter physics.
In the basic `electron-gas' models of metals, the {\em extended} (generalized)  eigenfunctions corresponding to the ac component play the main role in  conduction.  
The last  statement needs to be qualified. Certain conduction issues, such as 
adiabatic transport and the celebrated quantum hall effect, also require a good  
understanding of the implications of pure-point 
spectra~\cite{H82,BESB,AvS2,Aiz_Graf}. Moreover, in the absence of ac spectrum, 
the divergence of the localization length, even at a single energy, plays an important role in the dynamics generated by $H_\lambda (\omega)$~\cite{Sims_Stolz}.  
The previous examples show that  the relation of spectral and transport properties can be subtle. 
However,  
as was noted by Miller and Derrida~\cite{MilDer93}, 
the following simple hypothetical experiment illustrates the correspondence of ac spectra with conduction. 

 \begin{figure}[h]
\begingroup\makeatletter%
\gdef\SetFigFont#1#2#3#4#5{%
  \reset@font
  \fontsize{10}{#2pt} % 
  \fontfamily{#3}\fontseries{#4}\fontshape{#5}%
  \selectfont}%
\endgroup%
\begin{center}\begin{picture}(0,0)%
\includegraphics{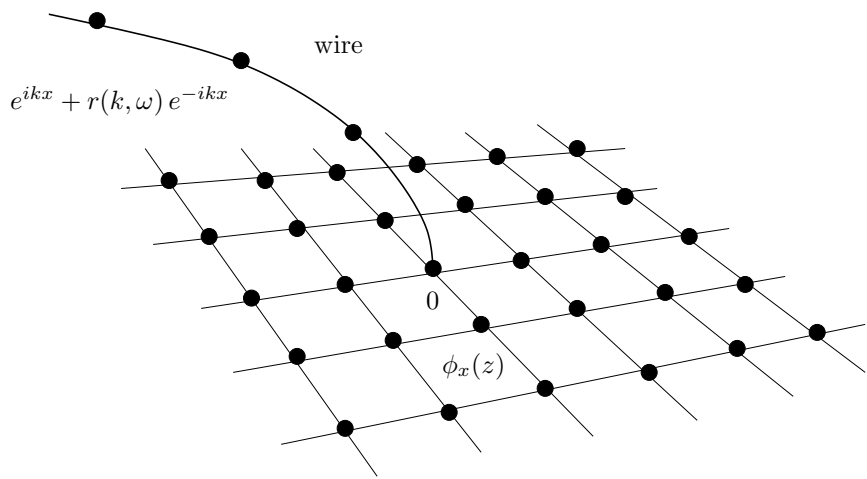}%
\end{picture}%
\setlength{\unitlength}{2644sp}%
\begingroup\makeatletter\ifx\SetFigFont\undefined%
\gdef\SetFigFont#1#2#3#4#5{%
  \reset@font\fontsize{#1}{#2pt}%
  \fontfamily{#3}\fontseries{#4}\fontshape{#5}%
  \selectfont}%
\fi\endgroup%
\begin{picture}(8112,4370)(2776,-8173)
\put(5626,-4186){\makebox(0,0)[lb]{\smash{{\SetFigFont{9}{10.8}{\rmdefault}{\mddefault}{\updefault}{wire}%
}}}}
\put(6676,-6586){\makebox(0,0)[lb]{\smash{{\SetFigFont{9}{10.8}{\rmdefault}{\mddefault}{\updefault}{$0$}%
}}}}
\put(2776,-4711){\makebox(0,0)[lb]{\smash{{\SetFigFont{9}{10.8}{\rmdefault}{\mddefault}{\updefault}{$e^{ikx}+r(k,\omega)\, e^{-ikx}$}%
}}}}
\put(6826,-7186){\makebox(0,0)[lb]{\smash{{\SetFigFont{9}{10.8}{\rmdefault}{\mddefault}{\updefault}{$\phi_x(z)$}%
}}}}
\end{picture}%

\caption{The setup for a scattering experiment.  Current is sent down a wire  which is attached to  the graph.}\label{graph2}
\end{center}
\end{figure}
Consider attaching an external wire at a vertex $0$ and sending down it particles  to probe the graph; as depicted  in  Figure~\ref{graph2}.  
For simplicity, assume that the dynamics of the particle is generated by an operator which, within the graph,  is given by $H_\lambda (\omega)$, 
along the wire is given by the one-dimensional operator $-\Delta + C$ with some constant potential  $C$, and the contact 
is described through a suitable local rule.  
When particles are sent down at  energy $E$ and decay rate $\eta>0$ the steady state amplitude for observing a particle at $x$ is given by
\begin{equation}
\psi_x(z) = \left\{ \begin{array}{ll} e^{ikx} + r(k;\omega)\, e^{-ikx} &
    \mbox{along the wire} \\ \phi_x(z) & \mbox{in the graph}, \end{array} \right.  
    \end{equation}
where $z  = E+i \eta = 4 \sin^2(k/2) + C$ 
and $\phi(z)$ is an $\ell^2$-solution of the Schr\"odinger equation $\left( -\Delta +
  \lambda \, V(\omega) - z \right) \phi = {\rm const.}\; \delta_0$ in the graph. 
This solution is unique up to a constant factor.  Here $r(k;\omega)$ is the reflection coefficient.  Its value is readily determined from the resolvent, and one finds:   
\begin{equation}
|r(k;\omega)|<1  \qquad   \Leftrightarrow \qquad 
  \, {\rm Im}
    \left\langle \delta_0, \frac{1}{H_{\lambda}(\omega) -E - i0} \; \delta_0
    \right\rangle  \  > \   0  \, .
\end{equation}
The graph absorbs the current of particles and conducts it to infinity if
$|r(k;\omega)|<1 $, and we see that this happens precisely 
when there is an ac component in the spectral decomposition of $\delta_0$, as its density is given by the imaginary part of 
the diagonal element of the Green function  times $\pi^{-1}$. \\
 
%%%%%%%%%%%% 

The remaining component of the spectral measure  
is referred to as the 
singular part. In a phenomenon known as Anderson localization \cite{Anderson}, this part almost surely includes  a pure-point component which   consists of a countable collection of eigenvalues 
$\{E_{n}( \omega)\}$  forming a dense subset of the non-random set $\Sigma_{\rm pp}$.  The corresponding 
eigenfunctions are exponentially localized, satisfying: 
\begin{equation}  \label{eq:loc}
|\psi_{n; x}(\omega)| \ \le \ A_{n}(\omega) \; e^{-|x-x_{n}( \omega)|/\xi} \, .
\end{equation} 
Random Schr\"odinger operators  have been the subject of intensive mathematical research \cite{CaLa90,PF,Stoll}.  However, most of the results which were obtained pertain to the regime of Anderson localization, with the above characteristics. 
 
\begin{figure}[hbt]
\begingroup\makeatletter%
\gdef\SetFigFont#1#2#3#4#5{%
  \reset@font
  \fontsize{10}{#2pt}% 
  \fontfamily{#3}\fontseries{#4}\fontshape{#5}%
  \selectfont}%
\endgroup%
\begin{center}
\begin{picture}(0,0)%
\includegraphics{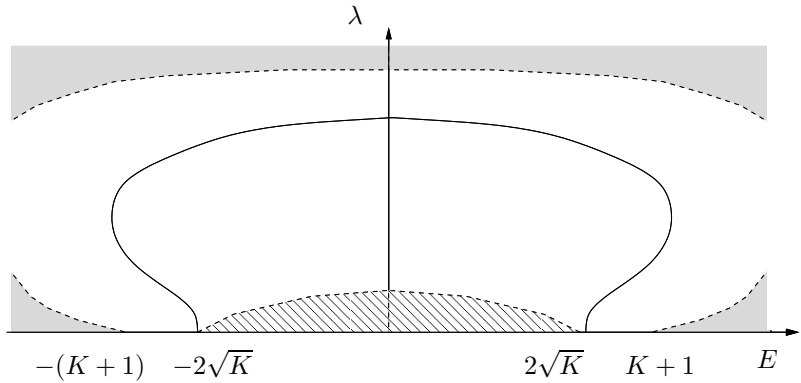}%
\end{picture}%
\setlength{\unitlength}{1973sp}%
\begingroup\makeatletter\ifx\SetFigFont\undefined%
\gdef\SetFigFont#1#2#3#4#5{%
  \reset@font\fontsize{#1}{#2pt}%
  \fontfamily{#3}\fontseries{#4}\fontshape{#5}%
  \selectfont}%
\fi\endgroup%
\begin{picture}(0,0)%
\includegraphics{bomb.eps}%
\end{picture}%
\setlength{\unitlength}{1973sp}%
\begingroup\makeatletter\ifx\SetFigFont\undefined%
\gdef\SetFigFont#1#2#3#4#5{%
  \reset@font\fontsize{#1}{#2pt}%
  \fontfamily{#3}\fontseries{#4}\fontshape{#5}%
  \selectfont}%
\fi\endgroup%
\begin{picture}(10004,4430)(1179,-3644)
\put(5661,814){\makebox(0,0)[lb]{\smash{{\SetFigFont{6}{7.2}{\rmdefault}{\mddefault}{\updefault}{$\lambda$}%
}}}}
\put(10801,-3511){\makebox(0,0)[lb]{\smash{{\SetFigFont{6}{7.2}{\rmdefault}{\mddefault}{\updefault}{$E$}%
}}}}
\put(1726,-3586){\makebox(0,0)[lb]{\smash{{\SetFigFont{6}{7.2}{\rmdefault}{\mddefault}{\updefault}{$-(K+1)$}%
}}}}
\put(3351,-3586){\makebox(0,0)[lb]{\smash{{\SetFigFont{6}{7.2}{\rmdefault}{\mddefault}{\updefault}{$\, \,  -2\sqrt{K}$}%
}}}}
\put(7876,-3586){\makebox(0,0)[lb]{\smash{{\SetFigFont{6}{7.2}{\rmdefault}{\mddefault}{\updefault}{$2\sqrt{K}$}%
}}}}
\put(9151,-3586){\makebox(0,0)[lb]{\smash{{\SetFigFont{6}{7.2}{\rmdefault}{\mddefault}{\updefault}{$K+1$}%
}}}}
\end{picture}%
\caption{A schematic sketch of the `phase diagram' of the spectrum of the discrete Schr\"odinger operator on a homogeneous tree,  
$ H_\lambda(\omega) = T + \lambda V(\omega) $ 
with  iid   random potential of zero mean. The solid line indicates the expected mobility edge separating the ac and the pure-point spectral regimes.  The dotted lines delineate regions for which ac (lined) and pure point (shaded) spectra were established.   
It is conjectured that  finite dimensional  graphs have a similar phase diagram, however there is at present no proof of the existence of ac spectrum in any example of a local operator with homogeneous disorder in finite dimensions.}     
\label{bomb}
\end{center}
\end{figure}

Localization  is expected to be more pronounced in low spatial dimensions.  Indeed, among the first rigorous results
was a proof \cite{GoMo77} that for a class of one dimensional random Schr\"odinger operators 
\begin{equation}\label{eq:1d}
	\sigma(H_\lambda (\omega)) = \Sigma_{\rm pp} 
\end{equation}
at any strength of the disorder, $\lambda \neq 0$. Later works confirmed that \eqref{eq:1d} 
is generic for one dimensional Schr\"odinger operators with extensive disorder; for a review of results see \cite{Sto02}.
In all dimensions Anderson localization is  present for large disorder or 
energies outside the spectrum of the zero-disorder counterpart of 
the operator.  This has been established rigorously for various models using either 
the multiscale-approach \cite{FrSp83,FrMa85,DrKl89,GeKl01} 
or the fractional moment method \cite{AM,A2,AENSS}. 
Aside from that, a common prediction is that 
part of the ac spectrum should be stable under 
weak disorder provided the spatial dimension is greater than two.   So far, this question has been largely untouched by the rigorous works. 
The only existing results are for   
the ``infinite dimensional'' case of  tree operators 
\cite{Klein95,Klein96,Klein98,ASW05,ASW05b,AW}, as depicted in Figure~\ref{bomb}.\\

The purpose of this short note is to summarize the essentials of \cite{ASW05,ASW05b,AW} where the tree scenario is revisited and a new method for 
establishing the stability of the ac spectrum for tree operators is presented.  In particular, we will review the fluctuation based argument which is at the core of that method.

%%%%%%%%%%%%%%%%%%%%%%%%%%%%%%%%%%%%%%%%%%%%%%%%%%%%%%%%%%%%%%%%%%%%%%%%%%%%
%%%%%%%%%%%%%%%%%%%%%%%%%%%%%%%%%%%%%%%%%%%%%%%%%%%%%%%%%%%%%%%%%%%%%%%%%
\section{Discrete Schr\"odinger operator on a regular rooted tree}

Let $\mathbb{T}$ be the set of vertices of a regular rooted tree with branching number $K \geq 2$.   The tree version of the discrete random Schr\"odinger operator, acting in $\ell^2(\mathbb{T})$, is 
\begin{equation} \label{eq:dism}
H_{\lambda}( \omega) = T + U + \lambda \, V(\omega)
\end{equation}
where $T$ is the adjacency matrix, $U $ is a potential which is radial and
periodic in the distance to the root, and $V(\omega)$ a random potential given by a sequence
$\omega = \{ \omega_x \}_{x \in \mathbb{T}}$  of random variables.  The parameter $\lambda \in
\mathbb{R}$ controls here the strength of the disorder. 
This version of the discrete random Schr\"odinger operator  has attracted attention early on, as an instructive example for the study of Anderson localization~\cite{Abou73,Abou74}.  \\  

Let us first note what is known 
about the spectral properties of the background operator corresponding to $ \lambda = 0 $.
Here the following observation, which we take from \cite{ASW05}, is useful.
\begin{lemma}\label{lem:1d} 
If $\{V_x\}_{x \in \mathbb{T}} $ is a radial function on $ \mathbb{T} $, then the ac spectrum of $ H = T + V $ on $ \ell^2(\mathbb{T}) $
coincides with the scaled up version of the ac spectrum of $ h^+ = T + V / \sqrt{K} $ on $ \ell^2(\mathbb{N}_0) $, satisfying:
\begin{equation}\label{eq:t1d}
 \sigma_{\rm ac} \left( H \right) \ = \  \sqrt{K} \; \sigma_{\rm ac} \left( h^+ \right) \, .
\end{equation}
For the singular spectrum an inclusion holds, $
   \sigma_{\rm sing} \left( H \right) \ \supset \
 \sqrt{K} \; \sigma_{\rm sing} \left( h^+ \right) $.
\end{lemma}

This statement has the following consequences for radial tree operators:  
\begin{enumerate}
\item[i.]  The ac spectrum of  $H_0 = T +U $ on $\ell^2(\mathbb{T})$ is non-empty, 
$ \sigma_{\rm ac}(H_0) \neq \emptyset$,  
and, as with periodic potentials,  has a band structure. An example is 
$\sigma_{\rm ac}(T) = [ -2 \sqrt{K}, 2
\sqrt{K} ]$, in which case there are no gaps.

\item[ii.]  The ac spectrum is unstable under perturbations by arbitrarily weak but radially symmetric disorder of  homogeneous strength.  For the random operator defined by setting
$\omega_x = \xi_{|x|}$, with $\{ \xi_n \}_{n \in \mathbb{N}_0}$ a
sequence of iid non-constant random variables, at any 
$\lambda \neq 0$:  
\begin{equation}
\sigma_{\rm ac} \left( H_{\lambda}( \omega) \right) = \emptyset
\end{equation} 
for almost every $\omega$.  This is due to the well known effect of disorder in one dimension. 
\end{enumerate}

The last comment is to be contrasted with the main result presented below.  As will be explained, it shows that  the ac spectra are stable under weak disorder which is either uncorrelated, or meets the following weak-correlation condition. 

\begin{definition} The random variables $ \{ \omega_x
  \}_{x \in \mathbb{T}}$ are said to be {\em weakly correlated} iff there exists a $\kappa \in (0,1]$ such that 
  \begin{figure}[h]
\begingroup\makeatletter%
\gdef\SetFigFont#1#2#3#4#5{%
  \reset@font
  \fontsize{10}{#2pt}% 
  \fontfamily{#3}\fontseries{#4}\fontshape{#5}%
  \selectfont}%
\endgroup%
\begin{center}
\begin{picture}(0,0)%
\includegraphics{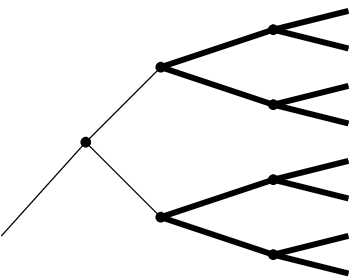}%
\end{picture}%
\setlength{\unitlength}{2368sp}%
\begingroup\makeatletter\ifx\SetFigFont\undefined%
\gdef\SetFigFont#1#2#3#4#5{%
  \reset@font\fontsize{#1}{#2pt}%
  \fontfamily{#3}\fontseries{#4}\fontshape{#5}%
  \selectfont}%
\fi\endgroup%
\begin{picture}(2937,2188)(2314,-2655)
\put(3201,-961){\makebox(0,0)[lb]{\smash{\SetFigFont{7}{8.4}{\rmdefault}{\mddefault}{\updefault}{$x$}%
}}}
\put(3201,-2311){\makebox(0,0)[lb]{\smash{\SetFigFont{7}{8.4}{\rmdefault}{\mddefault}{\updefault}{$y$}%
}}}
\put(5251,-961){\makebox(0,0)[lb]{\smash{\SetFigFont{7}{8.4}{\rmdefault}{\mddefault}{\updefault}{$\mathbb{T}_x$}%
}}}
\put(5251,-2161){\makebox(0,0)[lb]{\smash{\SetFigFont{7}{8.4}{\rmdefault}{\mddefault}{\updefault}{$\mathbb{T}_y$}%
}}}
\end{picture}

\end{center}
\end{figure}
  for any $x \neq
  y \in \mathbb{T}$ which are common forward neighbors to some vertex and any pair of functions $X, Y : \mathbb{R}^{ \mathbb{T}} \to [0, \infty)$
  measurable with respect to the $\sigma$-algebra associated with the forward subtrees
  $\mathbb{T}_x$ and $\mathbb{T}_y$ respectively, one has
\begin{equation}
\mathbb{E} \left[ X( \omega) Y( \omega) \right] \geq \kappa \; \mathbb{E}
\left[ X( \omega) \right] \mathbb{E} \left[ Y( \omega) \right].
\end{equation}
\end{definition}

Any iid sequence forms a trivial example of such weakly
correlated random variables.  A broader class of examples can be found in \cite{ASW05}. \\

The main result of \cite{ASW05} reads as follows:

\begin{theorem}\label{thm:ctden} Let $\{ \omega_x \}_{x \in \mathbb{T}}$ be stationary,
  weakly correlated, and suppose that $\mathbb{E} \left[ \log \left( 1 + |
    \omega_0 | \right) \right] < \infty$. Then, for every $I \subset
\sigma_{\rm ac}(H_0)$ with $I \cap \sigma_{\rm sing}(H_0) =
\emptyset$:  
\begin{multline}\label{eq:maindisc}
\lim_{\lambda \to 0} \mathbb{E} \left[ \, \int_I  \left| \, {\rm Im}
    \left\langle \delta_0, \frac{1}{H_{\lambda}(\omega) -E - i0} \; \delta_0
    \right\rangle - {\rm Im} \left\langle \delta_0, \frac{1}{H_0 -E - i0} \, \delta_0
    \right\rangle \right| \, dE \, \right] \\ = 0,
\end{multline}
\end{theorem}
Some remarks are in order:
\begin{enumerate}
\item[i.] Eq.~(\ref{eq:maindisc})  is a statement of 
  $L^1$-convergence of the ac densities, which of course implies weak convergence of the ac components of the spectral measures. 
  As a consequence of \eqref{eq:maindisc} the ac spectrum is stable in Lebesgue sense, i.e., for all Borel sets $I \subset \mathbb{R}$
\begin{equation} \label{eq:stac}
\lim_{\lambda \to 0} \left| \, \sigma_{\rm ac} \left( H_{\lambda}(
    \omega) \right) \cap I \right| = \left| \, \sigma_{\rm ac}(H_0) \cap I \right| \, ,
\end{equation}
where $ | \cdot | $ denotes the Lebesgue measure.
\item[ii.] In the case of $U=0$ and $\{ \omega_x \}_{x \in
  \mathbb{T}}$ iid, more is known. Klein \cite{Klein95,Klein98} proved that  for every $ 0 < |E| < 2 \sqrt{K}  $
there exists $ \lambda(E) > 0 $ such that 
\begin{equation}
\sup_{ \substack{ 0 < \eta \leq 1, \ \ 0< |E'| \leq E \\ |\lambda| \leq \lambda(E)} }\mathbb{E} \left[ \left| 
\left\langle \delta_0, \frac{1}{H_{\lambda}(\omega) -E' - i \eta} \, \delta_0
    \right\rangle \right|^2 \right] < \infty,
\end{equation}
which is sufficient to ensure that the spectrum in this interval is
purely absolutely continuous.  

\item[iii.] For the last example ($U=0$)
pure point spectrum was proven at large disorder and also for small disorder and energies $|E| \geq K+1$ in \cite{A2}. 
This leaves still unresolved the issue of band edge localization in the energy regime $ |E | \in [2 \sqrt{K}, K+1] $. 
\end{enumerate}   

\section{Radial quasi-periodic operators}

Theorem~\ref{thm:ctden} has been extended in \cite{AW} to certain cases where $ \{ U_x \}_{x \in \mathbb{T}} $ 
is a radial and quasi-periodic (qp) function, i.e.,
\begin{equation} \label{eq:qp}
U_x(\theta) = u(S^{|x|} \theta ) \, ,
\end{equation} 
where $ u: \Xi \to \mathbb{R} $ is a continuous function on a multidimensional torus $ \Xi $ on which 
$ S $ acts as an ergodic shift, $ S \theta = (\theta + 2\pi \alpha ) \mod 2\pi $, with a frequency vector $ \alpha $.
The stability results then refers to the operator
\begin{equation} 
	H_\lambda(\theta,\omega) = T + U(\theta) + \lambda \, V(\omega) 
\end{equation}
on $ \ell^2(\mathbb{T}) $.   The operator $ H_0(\theta) $ represents a `fanned out' version of the qp one-dimensional operator  
\begin{equation}\label{eq;h}
h(\theta) = T + u(S^{|x|} \theta )/\sqrt{K}
\end{equation} 
on $ \ell^2(\mathbb{Z}) $.  The precise assumptions on 
$ u $ are formulated in terms of the extensively studied properties of $h(\theta)$.

For a variety of one-dimensional qp operators $ h(\theta) $ it has been established by KAM and duality methods that they
admit pairs of {\em Bloch-Floquet eigenstates} for almost
all energies in its ac spectrum \cite{DiSi75,BLT83,Sin87,CD89,Eli92,Puig04,GJLS97,Jit99,BJ02,Puig}.  The latter are defined as follows.
\begin{definition}
The operator $ h(\theta) $ is said to admit a \emph{Bloch-Floquet eigenstate} with energy $ E $ and quasi-momentum $ k \in (-\pi,\pi] $ 
iff there is a non-vanishing function $ \psi: \mathbb{Z} \times \Xi \to \mathbb{C} $ such that 
\begin{align}
	(i) \quad & h(\theta) \psi(\theta) = E \, \psi(\theta) && \qquad \mbox{in the weak sense} \notag \\
	(ii) \quad & \psi_x(\theta) = \exp(ikx) \, \varphi(S^x\theta) && \qquad \mbox{for some continuous $ \varphi: \Xi \to \mathbb{C} $.} \notag
\end{align}
\end{definition}
Bloch-Floquet eigenstates can be viewed as states which are covariant functions on the operator, with respect to translations.  They naturally come in pairs: if $ \psi(\theta) $ is one, then its complex conjugate is also one with the same energy, but reversed quasi-momentum.  
While not true for all~\cite{Las93},  
many  qp operators $ h(\theta) $ admit pairs of linearly independent, complex-conjugate Bloch-Floquet eigenstates throughout their 
ac spectra (with the exclusion of countable collections of energies). The most prominent example is the almost Mathieu operator 
which corresponds to $ u(\theta) = u_0 \, \cos(\theta) $ with $ \theta \in [0,2\pi) $.\\

Within this setup, the main result of \cite{AW} is: 
 
\begin{theorem} Let $ \{\omega_x\}_{x\in \mathbb{T}} $ be as in Theorem~\ref{thm:ctden}. Suppose that for almost every 
$ E \in \sigma_{\rm ac}(h(\theta)) $ the operator
$h(\theta) $ admits a  pair of linearly independent, complex-conjugate Bloch-Floquet eigenstates. 
Then, in the weak disorder limit: 
\begin{equation}
\lim_{\lambda \to 0} \left| \, \sigma_{\rm ac} \left( H_{\lambda}(\theta, \omega) \right) \cap I \right| 
	= \left| \, \sigma_{\rm ac}(H_0(\theta)) \cap I \right| \, .
\end{equation}
\end{theorem}
The r\^ole played by the Bloch-Floquet eigenstates in the proof will be briefly commented upon at the end of Section~\ref{sec:proof}.

%%%%%%%%%%%%%%%%%%%%%%%%%%%%%%%%%%%%%%%%%%%%%%%%%%%%%%%%%%%%%
%%%%%%%%%%%%%%%%%%%%%%%%%%%%%%%%%%%%%%%%%%%%%%%%%%%%%%%%%%%%
\section{Random quantum trees}\label{sec:qg}    

In this section we consider the Laplacian on the Hilbert space 
$\bigoplus_{e \in E}L^2(0,L_e)$ where $E$ is
the set of edges corresponding to $\mathbb{T}$. For any edge $e \in
E$ the length of $e$ is denoted by $L_e$ with orientation (in
particular, the direction of increase) taken away from the root. Any
function $\psi $ in the above Hilbert space may be written as $\psi = \{ \psi_e \}_{e
  \in E}$, and the quantum tree graph Laplacian acts as a second derivative 
\begin{equation} \label{eq:qgm}
 \Delta_{\mathbb{T}} \psi  = \{  \psi_e'' \}_{e \in E}
\end{equation}
on the domain of functions $\psi$ for which each $\psi_e \in H^2(0,
L_e)$ subject to certain boundary conditions, which we take as follows:
\begin{itemize}
\item[i.] at the root: 
\begin{equation}
\cos( \alpha) \psi(0) - \sin( \alpha) \psi'(0) = 0,  
\end{equation}
for some fixed $\alpha \in [0, \pi)$. 
\item[ii.] at internal vertices the functions are required to satisfy the Kirchoff/Neumann conditions, i.e., $\psi$ is continuous and
\begin{equation}\label{Kirch}
\psi_e'(L_e) = \sum_{e' \in \mathcal{N}_e^+} \psi_e'(0).
\end{equation} 
\end{itemize}
These conditions guarantee that $\Delta_{\mathbb{T}}$ is self-adjoint
\cite{KosSch99} and that $\sigma_{\rm ac}( - \Delta_{\mathbb{T}})$ is independent
of $\alpha$. 

For the regular tree, i.e., $L_e=L$ for all $e \in E$, the spectrum is explicitly known \cite{SoSo02,Sol04},
\begin{equation}
\sigma_{\rm ac}( - \Delta_{\mathbb{T}}) = \bigcup_{n=0}^{ \infty}
\left[  \left( \frac{ \pi (n+1) + \theta}{L} \right)^2,  \left( \frac{ \pi n + \theta}{L} \right)^2\right],
\end{equation}
where $\theta = \mbox{arctan} \frac{K-1}{2 \sqrt{K}}$. Moreover, there occur infinitely degenerate eigenvalues in the 
band gaps at the Dirichlet
eigenvalues of the Laplacian on an interval of length $ L $.\\

Our aim is to study random deformations
$\mathbb{T}_{\lambda}(\omega)$ of such a regular tree graph. 
We keep the vertex set fixed but define
random edge lengths in terms of 
\begin{equation}
	L_e(\omega) := L \, \exp( \lambda \omega_e)
\end{equation} 
where
$\omega = \{ \omega_e \}_{e \in E}$ are bounded iid random variables,
and $\lambda \in [0,1]$ is the disorder parameter. 
For this setup the main result of \cite{ASW05} reads as follows.
\begin{theorem}\label{thm:qg}
For any Borel set $I \subset \mathbb{R}$ and almost every
  $\omega$,
\begin{equation}
 \lim_{\lambda \to 0} \left|\, \sigma_{\rm ac} \left( 
     -\Delta_{\mathbb{T}_{\lambda}( \omega)} \right) \cap I \right| =
 \left| \, \sigma_{\rm ac}( -\Delta_{\mathbb{T}}) \cap I \right|.
\end{equation}
\end{theorem}

Let us close this section with few remarks: 

\begin{itemize}
\item[i.] The result of Theorem~\ref{thm:qg} remains true if $\{ \omega_e
\}_{e \in E}$ are weakly correlated and it is expected to be false if $\{ \omega_e \}_{e \in E}$ is radial. The results are also valid 
if one replaces \eqref{Kirch} by some $\alpha$-type boundary conditions - provided $\alpha$ is either constant or a radial function of $x$.
\item[ii.] There are only a few published results on localization for quantum graphs.  These are generally expected to be similar to those of the discrete Schr\"o\-dinger operators.  However, this setup provides a natural context for the study of other issues. The work \cite{SchSmil00}  presents a localization result 
for one-dimensional quantum graphs  
derived through the analysis of the interference of periodic orbits in the presence of random scatterers.  Better understanding of such cancellations is of independent interest with relation to trace formulae and issues of dynamics in the presence of disorder.  
\end{itemize}

%%%%%%%%%%%%%%%%%%%%%%%%%%%%%%%%%%%%%%%%%%%%%%%%%%%%%%%%%%%%%%%%%%%%%%
%%%%%%%%%%%%%%%%%%%%%%%%%%%%%%%%%%%%%%%%%%%%%%%%%
\section{Sketch of the proof}\label{sec:proof}

In the remainder, we will sketch the 
proof of the stability of the ac spectrum for the random tree
models which were introduced above. For simplicity, we focus on the discrete model 
(\ref{eq:dism}) with $U=0$.  As is shown in \cite{ASW05b}, a similar 
analysis applies to the quantum graph model, once one has identified the appropriate quantities.

Our analysis begins by identifying a particular solution, familiar from the study of
one dimensional Sturm-Liouville equations, which we refer
to as the Weyl-Titchmarsh function. This function can be
understood from a scattering theoretic perspective, and within this
context, one is naturally led to certain observations concerning the
conservation (or loss, when $\eta \neq 0$)  of current  along the edges of the tree. Analysis of the current conservation inequality
provides a bound on the fluctuations of the random Green  functions 
in terms of a quantity which we describe as a Lyapunov exponent on
the tree. 
Similar to the one dimensional theory, this Lyapunov
exponent, which is the negative of the real part of a Herglotz
function, vanishes on the ac spectrum of the background
operator. Quite generally, based on arguments which are natural for functions in $H^p$ spaces, 
averages of such quantities are 
continuous in their parameters, and from this fact, in particular,
(\ref{eq:jctle}), we can prove that
the distribution of these random Green's functions collapse to a
unique point in the limit of weak disorder. Uniqueness of the solution of the 
underlying recursion relation identifies this point as the
free Green's function. Theorem~\ref{thm:ctden} 
follows from these observations, although some attention to the details is necessary.
%%%%%%%%%%%%%%%%%%%%%%%%%%%%%%%%%%%%%%%%%%%%%%%%%%%%
%%%%%%%%%%%%%%%%%%%%%%%%%%%%%%%%%%%%%%%%%%%%%%%%%%%

\subsection{The Weyl-Titchmarsh function on the tree}
Consider the scattering experiment depicted in Figure~\ref{graph2} in case $ \mathbb{Z}^d $ replaced by $ \mathbb{T} $ and the wire is attached to the root of $ \mathbb{T} $.
Let 
$\psi(z)$ be an $\ell^2$-solution of the Schr\"odinger inside the tree, which is  unique up to a constant factor and vanishes nowhere.
Therefore the following quantity, which we refer to as the Weyl-Titchmarch function, is unique and well defined.  

\begin{definition} For any $x \in \mathbb{T}$ and $z \in \mathbb{C}$ take
\begin{equation}\label{eq:WT}
\Gamma_x(z) := - \frac{ \psi_x(z)}{ \psi_{x^-}(z)},
\end{equation}
where $x^-$ is the backward neighbor of $x$.
\end{definition}

The function $\Gamma_x$ has several interesting properties:

\begin{itemize}
\item[i.] Since  $ \psi(z)$ satisfies the Schr\"odinger equation, $ \Gamma(z) $ satisfies the following {\em recursion relation}
\begin{equation}
\Gamma_x(z) = \frac{1}{ \lambda \omega_x - z - \sum_{y \in
    \mathcal{N}_x^+} \Gamma_y(z)},
\end{equation}
where $ \mathcal{N}_x^+ $ is the set of forward neighbors of $ x $.
\item[ii.] One may rewrite $ \Gamma_x(z) $ in terms of the resolvent of $ H_\lambda(\omega) $ restricted to the Hilbert space over the 
forward subtree $ \mathbb{T}_x $ 
starting at $ x $,
\begin{equation}
\Gamma_x(z) = \left\langle \delta_x, \frac{1}{H_\lambda(\omega) |_{\mathbb{T}_x} - z } \delta_x \right\rangle.
\end{equation}
Therefore,
 $\Gamma_x$ is a {\em Herglotz-Nevanlinna} function, i.e.,
$\Gamma_x$ is analytic in $\mathbb{C}^+$ and $ {\rm Im} \,\Gamma_x(z) >0$ if ${\rm Im}z>0$.
\item[iii.] The {\em ac-density} of the Schr\"odinger operator $H_\lambda(\omega) $ associated with the vector $ \delta_0 $ at the root
is given by $ {\rm Im} \,\Gamma_0(E+i0)/\pi $.
\item[iv.] The {\em current} carried by $\psi(z)$ along the bond
from $x$ to $y$ is
\begin{equation} \label{eq:cur}
J_{xy}(z) = | \psi_x(z)|^2 \, {\rm Im}\, \Gamma_y(z),
\end{equation} 
and therefore the {\em current conservation / loss} at each vertex may
be written as
\begin{equation} \label{eq:curcon}
J_{x^-x}(z) \geq \sum_{y \in \mathcal{N}_x^+} J_{xy}(z),
\end{equation} 
with equality for real $z$.
\end{itemize}

Rewriting (\ref{eq:curcon}) in terms of (\ref{eq:cur}), taking the
logarithm, and then averaging yields
\begin{equation} \label{eq:logbd}
\mathbb{E} \left[ \log \frac{1}{K} \sum_{y \in \mathcal{N}_x^+} {\rm
      Im}\, \Gamma_y \right]  -\mathbb{E} \left[ \log {\rm Im} \Gamma_x
  \right] \  \leq \  \mathbb{E} \left[ \log \frac{1}{K} \left| \frac{
        \psi_{x-}}{ \psi_x} \right|^2 \right] =: 2 \, \gamma_\lambda
\end{equation}
The quantity right side
of (\ref{eq:logbd}) may be identified as twice a {\em Lyapunov exponent} on the tree, 
since it measures the average decay rate of $ \psi $ in one generation.  
A standard application of Jensen's inequality to the left side
of (\ref{eq:logbd}) demonstrates that this quantity is
non-negative. As the Jensen inequality is strict unless the random
variables (over which one averages) cease to fluctuate, we are
naturally led to the core of our argument, which will be presented in the next subsection. 

%%%%%%%%%%%%%%%%%%%%%%%%%%%%%%%%%%%%%%%%%%%%%%%%%%%%%%%%%%%%
%%%%%%%%%%%%%%%%%%%%%%%%%%%%%%%%%%%%%%%%%%%%%%%%%%%%%%%%%%%
\subsection{Fluctuation Bounds}
We will bound the fluctuations of the
random variables $ \Gamma_x$ in terms of the Lyapunov exponent 
introduced above. To this end, we require an improvement of the Jensen
inequality which quantifies the `boost' due to fluctuations.

\begin{lemma} Let $\{ X_j \}_{j=1}^K$ with $K \geq 2$ be positive,
  weakly correlated, identically distributed random variables. Then,
  for any $\alpha \in (0, 1/2]$,
\begin{equation}
\mathbb{E} \left[ \log \frac{1}{K} \sum_{j=1}^K X_j \right] \ \geq \  
\mathbb{E} \left[ \log X_1 \right] + \frac{ \alpha^2 \kappa }{4}\;
\delta(X_1, \alpha)^2,
\end{equation} 
with the relative $\alpha$-width of $X_1$ which is defined as
\begin{equation}
\delta(X_1, \alpha) := \frac{ \xi_+(X_1, \alpha) - \xi_-(X_1,
  \alpha)}{ \xi_+(X_1, \alpha)} \,  ,
\end{equation}
where $ \xi_-(X_1, \alpha) := \sup \{ \xi \, : \, \mathbb{P}(X_1<\xi) \leq \alpha \} $ and $ \xi_+(X_1, \alpha) 
:= \inf \{ \xi \, : \, \mathbb{P}(X_1>\xi) \geq \alpha \} $.
\end{lemma} 

Applying this result with $\{ X_j \}_{j=1}^K $ taken to be $\{ {\rm Im}\,
\Gamma_y \}_{y \in \mathcal{N}_x^+}$, we obtain from (\ref{eq:logbd}) the first claim in the following 
\begin{theorem} \label{thm:flbd}
For all $\lambda \in \mathbb{R}$, $z \in \mathbb{C}^+$, and $\alpha
\in (0, 1/2]$: 
\begin{eqnarray}   
\delta \left( {\rm Im} \Gamma_0( z), \alpha \right)^2  &  \leq &   
\frac{8}{ \kappa \alpha^2}\; \gamma_{\lambda}(z) \, , \label{eq:flu1}\\[5pt] 
\delta \left( | \Gamma_0( z)|^2 , \alpha \right)^2  & \leq  &  
\frac{32(K+1)^2}{ \kappa \alpha^2} \; \gamma_{\lambda}(z)\, .\label{eq:flu2}\\ 
\end{eqnarray}
\end{theorem}

Since we have bounded the fluctuations of $\Gamma_x$ in terms of
the Lyapunov exponent, it is important to note several of its properties.  
\begin{enumerate}
\item[i.]  
From (\ref{eq:logbd}), \eqref{eq:WT} and stationarity, we conclude that
\begin{equation}
\gamma_{\lambda}(z) = - \mathbb{E} \left[ \log \sqrt{K} | \Gamma_0(z) | \right] \, .
\end{equation}
Hence, $\gamma_{\lambda}$ is a non-negative harmonic function of
$z \in \mathbb{C}^+$.    
\item[ii.]  Moreover, by Kotani theory \cite{Kot85} and the relation \eqref{eq:t1d} of the ac spectrum of tree operators with radial potentials  with that of the corresponding  operator on the line, we have $\gamma_0(E+i0)=0$ for almost every $E \in \sigma_{\rm ac}(H_0)$. 
\item[iii.]    
For $ {\rm Im}  z > 0$,  $\gamma_{\lambda}(z)$ is jointly continuous
in it's relevant parameters.  This yields the following continuity statement, which is derived through  the deformation of the  contour of integration  and the natural bounds on $\Gamma(z)$:   
\begin{equation} \label{eq:jctle}
\lim_{ \substack{ \lambda \to 0 \;\\ \eta \to 0^+}} \int_I
\gamma_{\lambda}(E+i \eta) \, dE = \int_I \gamma_0(E + i 0) \, dE = 0,
\end{equation}
for any Borel set  $I \subset \sigma_{\rm    ac}(H_0)$.    
\end{enumerate}  
\noindent The last relation means that, at least upon energy averaging, the right sides in \eqref{eq:flu1} and \eqref{eq:flu2} vanish in any joint limit 
$ \lambda, \eta \to 0 $. Accordingly, the relative fluctuations of $ {\rm Im} \Gamma_0 $ and $ | \Gamma_0 | $ vanish in this limit.

%%%%%%%%%%%%%%%%%%%%%%%%%%%%%%%%%%%%%%%%%%%%%%%%%%%%%%%%%%%%%%%%
%%%%%%%%%%%%%%%%%%%%%%%%%%%%%%%%%%%%%%%%%%%%%%%%%%%%%%%%%%%%%%%%%
\subsection{Putting it together}

To complete the result, we consider all possible distributional limits
of $\{ \Gamma_x \}_{x \in \mathbb{T}}$ as
$\lambda, \eta \to 0$, where the distribution $ \mathcal{D} $ refers to jointly the energy in some Borel subset of $ \sigma_{\rm ac}(H_0) $ 
and the randomness.  It should be understood there that $\Gamma_x$ initially depends  on   $\{\omega, \lambda, \eta, E\}$, and we are considering the possible   limits of these functions, for 
 $\lambda, \eta \to 0$.  
All such limits, or accumulation points, are characterized by
\begin{itemize} 
\item[i.] the vanishing distributional width of ${\rm
  Im}\Gamma_x$ and $| \Gamma_x|$.
\item[ii.] the stationarity:  $\Gamma_x \stackrel{
  \mathcal{D}}{=} \Gamma_y$ 
  \item[iii.] the limiting recursion relation
\begin{equation}\label{eq:recursion}
\Gamma_x \stackrel{ \mathcal{D}}{=} \frac{1}{-E - \sum_{y \in
    \mathcal{N}_x^+} \Gamma_y} \, , 
 \end{equation}
\end{itemize}
Since the (non-random) recursion relation $\Gamma =
\frac{1}{-E-K \Gamma}$ has a unique fixed point in $ \mathbb{C}^+ $, we conclude the distributional convergence
\begin{equation} 
\Gamma_x(E + i \eta) \stackrel{ \mathcal{D}}{\to }
\Gamma_x(E+i0),
\end{equation}
as $\lambda, \eta \to 0$.   Outside the singular spectrum $ \sigma_{\rm sing}(H_0) $ this implies the $ L^1 $-convergence claimed 
in \eqref{eq:maindisc}.\\

In case $H_0$ is radially periodic, with period $\tau$, the  stationarity of $\Gamma_x$ 
is replaced by periodicity, and in the recursion relation (\ref{eq:recursion})
the M\"obius transformation is replaced by its $\tau$-iterate.  

Let us finish with a remark on the issues showing up in this last step of the proof in case $ U $ is quasi-periodic as given by \eqref{eq:qp}
Here the above argument requires the uniqueness of solutions $ \Gamma $ with values in the upper half plane of the cocycle equation
\begin{equation}
	\Gamma(\theta) = \frac{1}{u(\theta) - E - K \, \Gamma(\theta)} 	
\end{equation} 
for almost all energies $ E \in  \sigma_{\rm ac}(H_0(\theta)) $. As was shown in \cite{AW},  this is the case if $ h(\theta) $ 
(given by \eqref{eq;h}) admits 
Bloch-Floquet eigenfunctions at Lebesgue-almost every energy in its  ac spectrum. The argument relies on a result in \cite{Puig} which proves the reducibility 
of the Schr\"odinger cocycle to a constant one under this assumption.  

\section{Some open challenges}  
It will be of interest to see progress on open issues related to the above discussion: 
\begin{itemize} 
\item[i.] Extend the analysis of stability of the ac spectra to finite dimensions.  
\item[ii.]  Clarify a possible relation of the spectral type with  the  level statistics.  
\end{itemize} 
The latter refers to the distribution of the energy gaps, resolved on the scale of the mean  energy spacing,   
for finite-volume  versions of the corresponding operators. 
Some partial results on the last point will be described elsewhere.  

\section*{Acknowledgements} 
The work described here has benefitted from our  enjoyable and stimulating discussions with Thomas Chen at Princeton Univ. and Uzy Smilansky at Weizmann Inst. Sci., where MA has enjoyed kind hospitality at  the Einstein Center for Theoretical Physics.  The work was supported in parts by the
  NSF grant PHY-9971149 (MA), an NSF Postdoctoral Fellowship (RS), and by the DFG grant Wa 1699/1 (SW).

%%%%%%%%%%%%%%%%%%%%%%%%%%%%%%%%%%%%%%%%%%%%%%%%%%%%%%%%%%%%%%%%%%%%%%%%
\bibliographystyle{amsalpha}

\end{document}